\documentclass[journal,onecolumn]{IEEEtran}
\usepackage[dvipdfmx]{graphicx}
\usepackage{color}

\ifCLASSINFOpdf
\else
\fi
\begin{document}
%
\title{High Spatial Resolution Neutron Transmission Imaging Using a Superconducting Two-Dimensional Detector}

\author{Hiroaki Shishido, Kazuma Nishimura, The Dang Vu, Kazuya Aizawa, Kenji M. Kojima, Tomio Koyama, Kenichi Oikawa, Masahide Harada, Takayuki Oku, Kazuhiko Soyama, Shigeyuki Miyajima, Mutsuo Hidaka, Soh Y. Suzuki, Manobu M. Tanaka, Shuichi Kawamata, and Takekazu Ishida
\thanks{H. Shishido is with Department of Physics and Electronics, Graduate School of Engineering, Osaka Prefecture University, Sakai, Osaka 599-8531, Japan, NanoSquare Research Institute, Osaka Prefecture University, Sakai, Osaka 599-8570, Japan, and The Center for Research \& Innovation in Electronic Functional Materials, Osaka Prefecture University, Sakai, Osaka 599-8570, Japan  (e-mail: Shishido@pe.osakafu-u.ac.jp)}
\thanks{K. Nishimura is with Department of Physics and Electronics, Graduate School of Engineering,Osaka Prefecture University, Sakai, Osaka 599-8531, Japan}
\thanks{T. D. Vu is with Division of Quantum and Radiation Engineering, Osaka
Prefecture University, Sakai, Osaka 599-8570, Japan, and with Materials and Life Science Division, J-PARC Center, Japan Atomic Energy Agency, Tokai, Ibaraki 319-1195, Japan}
\thanks{K. Aizawa, K. Oikawa, M. Harada, T. Oku and K. Soyama are with Materials and Life Science Division, J-PARC Center, Japan Atomic Energy Agency, Tokai, Ibaraki 319-1195, Japan}
\thanks{K. M. Kojima is with Center for Molecular and Materials Science, TRIUMF and Stewart Blusson Quantum Matter Institute, University of British Columbia, Vancouver, BC, V6T 2A3 and V6T 1Z4, Canada, and also with Division of Quantum and Radiation Engineering, Osaka Prefecture University, Sakai, Osaka 599-8570, Japan}
\thanks{T. Koyama is with Division of Quantum and Radiation Engineering, Osaka Prefecture University, Sakai, Osaka 599-8570, Japan}
\thanks{S. Kawamata and T. Ishida are with Division of Quantum and Radiation Engineering, Osaka Prefecture University, Sakai, Osaka 599-8570, Japan, and also with NanoSquare Research Institute, Osaka Prefecture University, Sakai, Osaka 599-8570, Japan}
\thanks{S. Miyajima is with Advanced ICT Research Institute, National Institute of Information and Communications Technology, 588-2 Iwaoka, Nishi-ku, Kobe, Hyogo 651-2492, Japan}
\thanks{M. Hidaka is with Advanced Industrial Science and Technology, Tsukuba, Ibaraki 305-8568, Japan}
\thanks{S. Y. Suzuki is with Computing Research Center, Applied Research Laboratory, High Energy Accelerator Research Organization (KEK), Tsukuba, Ibaraki 305-0801, Japan}
\thanks{M. M. Tanaka is with Institute of Particle and Nuclear Studies, High Energy Accelerator Research Organization (KEK), Tsukuba, Ibaraki 305-0801, Japan}
\thanks{Manuscript received month day, 202x; revised month day, 202x.}}

\markboth{Journal of IEEE transactions of Applied Superconductivity,~Vol.~, No.~, month~2021}
{Shell \MakeLowercase{\textit{et al.}}: Bare Demo of IEEEtran.cls for IEEE Journals}
%

\maketitle

\begin{abstract}
Neutron imaging is one of the most powerful tools for nondestructive inspection owing to the unique characteristics of neutron beams, such as high permeability for many heavy metals, high sensitivity for certain light elements, and isotope selectivity owing to a specific nuclear reaction between an isotope and neutrons.
In this study, we employed a superconducting detector, current-biased kinetic-inductance detector (CB-KID) for neutron imaging using a pulsed neutron source.
We employed the delay-line method, and high spatial resolution imaging with only four reading channels was achieved.
We also performed wavelength-resolved neutron imaging by the time-of-flight method for the pulsed neutron source.
We obtained the neutron transmission images of a Gd--Al alloy sample, inside which single crystals of GdAl$_3$ were grown, using the delay-line CB-KID.
Single crystals were well imaged, in both shapes and distributions, throughout the Al--Gd alloy.
We identified Gd nuclei via neutron transmissions that exhibited characteristic suppression above the neutron wavelength of 0.03\,nm.
In addition, the $^{155}$Gd resonance dip, a dip structure of the transmission caused by the nuclear reaction between an isotope and neutrons, was observed even when the number of events was summed over a limited area of $15\times12\mu$m$^2$.
Gd selective imaging was performed using the resonance dip of $^{155}$Gd, and it showed clear Gd distribution even with a limited neutron wavelength range of 1\,pm.
\end{abstract}


%
\IEEEpeerreviewmaketitle

\section{Introduction}
Superconducting detectors are used in several fields owing to their high sensitivity.
Transition edge sensors (TES) \cite{TES}, microwave kinetic conductance detectors (MKID) \cite{MKID_1, MKID_2}, and superconducting tunnel-junction detectors (STJ) \cite{STJ} are used in astronomy to detect electromagnetic waves from space.
Superconducting nanowire single-photon detectors and STJ are widely used for single-photon detection \cite{SNSPD, STJ_2}.
Besides these applications, superconducting detectors have been proposed for detecting other particle beams, including macromolecules \cite{STJ_m1} and neutrons \cite{STJ_n1, STJ_n2, TES_n}.
Both STJ- and TES-based superconducting neutron detectors have a neutron conversion layer containing atoms with large neutron absorption cross sections such as $^{10}$B and $^6$Li.

In our previous study, we developed a current-biased kinetic-inductance detector (CB-KID) as a superconducting neutron detector \cite{Shi15, Miy17, Shi18, Iiz19, Vu20}.
Employing a delay-line method, high spatial resolved neutron imaging with the resolution of 16.2\,$\mu$m was achieved with only four signal read terminals \cite{Iiz19}.
With a high temporal resolution of the delay-line CB-KID, high energy-resolved imaging was also achieved by the time-of-flight (TOF) method and pulsed neutron sources \cite{Shi18, Iiz19, Vu20}.
The device structure of CB-KID is similar to that of SNSPD: superconducting meanderlines are stacked on the superconducting ground plane, and DC bias current is passed through them.
However, the amplitudes of the bias current in the two methods are different.
Bias current of CB-KID is usually a few percentage of the critical current, whereas that of SNSPD is close to the critical current.
CB-KID detects a transient change in the density of the Cooper pair induced by dissipative irradiation through the transient change in kinetic inductance.
Therefore, the detection signal of CB-KID is proportional to the time derivative of the kinetic inductance, in contrast with MKID, which detects a change in the whole kinetic inductance.

The main advantage of the delay-line CB-KID is that it has both high spatial and temporal resolution;
thus, the energy dependence of neutron transmission can be observed with high spatial resolution using pulsed neutron sources.
Some isotopes have significant peaks in the energy dependence of the neutron total cross sections resulting from the resonance structure in the cross sections of the neutron-induced reaction \cite{Lam39}.
Therefore, a resonance dip, corresponding to peaks in the neutron total cross sections, appears in the neutron transmissions, and it indicates an isotope.
An elemental imaging technique based on resonance dip is called neutron resonance transmission imaging (NRTI) \cite{Fes15, Tre20}.
In addition to NRTI, the crystal structure, single crystallinity, and strains in a sample can be investigated by spatially-resolved neutron transmissions \cite{Iiz19}.
Although gadolinium-oxysulfide scintillator neutron detectors have achieved the highest spatial resolution of 2\,$\mu$m \cite{2um}, it may be difficult to perform energy-resolved imaging using a scintillator detector.
The combination of a $^{10}$B-doped microchannel plate and a Timepix readout provided a high spatial resolution energy-resolved neutron imager, and it was used to investigate neutron transmission imaging, tomography, NRTI, and strain mapping \cite{Tre20}.

In this study, we employed a delay-line CB-KID to perform neutron transmission imaging for a Gd--Al alloy sample containing GdAl$_3$ single crystals.
Single crystals of GdAl$_3$ were observed with the spatial resolution of at least 27.7 and 23.4\,$\mu$m for the $x$ and $y$ directions, respectively.
We represent a clear resonance dip of $^{155}$Gd even on a limited area of $15\times12\mu$m$^2$ and Gd selective NRTI result.

\section{Experimental method}
\subsection{Delay-line current-biased kinetic-inductance detector}
\begin{figure}
\begin{center}
\includegraphics[width=7cm, pagebox=cropbox, clip]{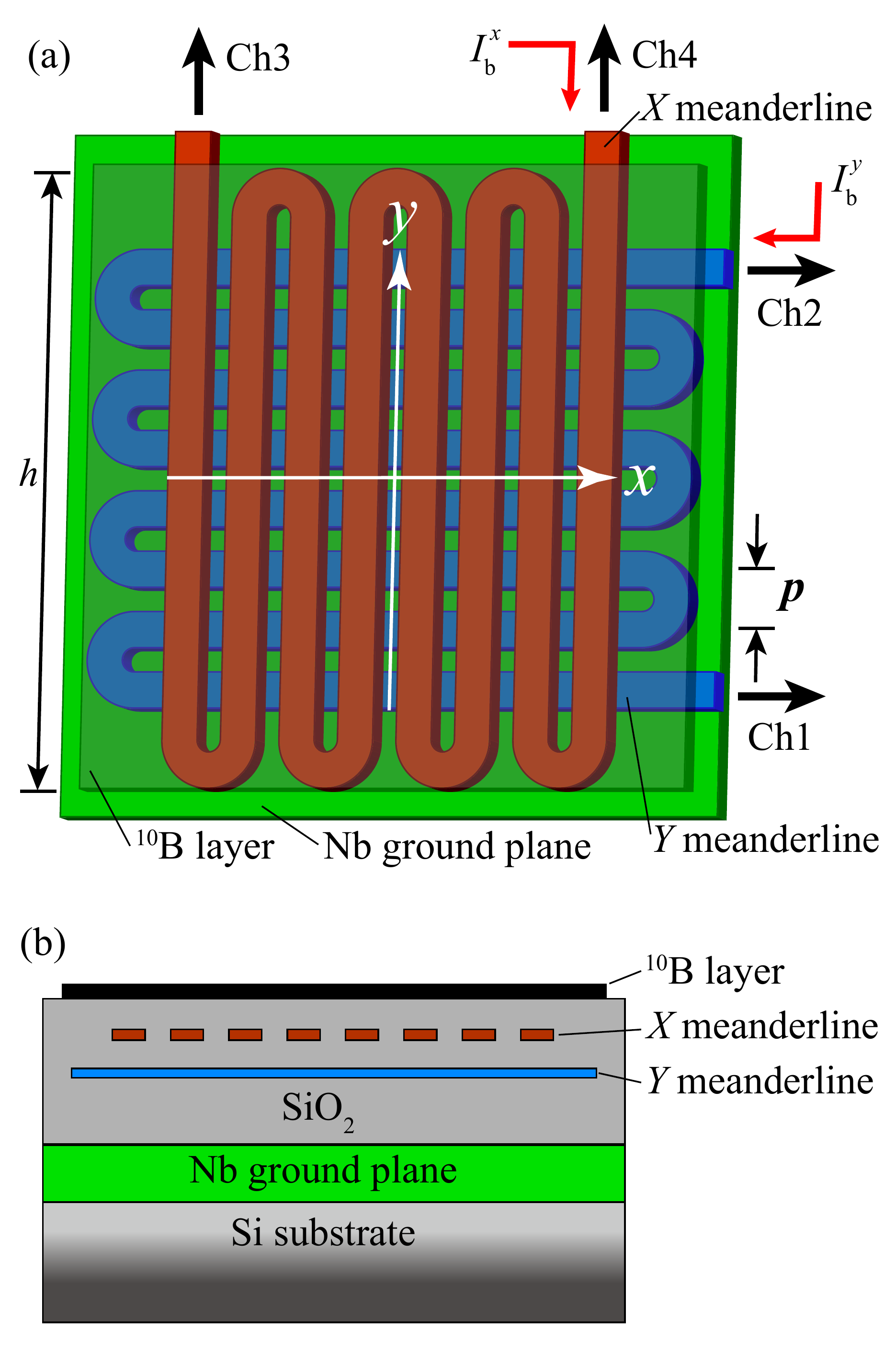}
\end{center}
\caption{\label{CBKID}Schematic of the CB-KID system---(a) it has stacked layers of a superconducting Nb ground plane, $X$ and $Y$ superconducting Nb meanderlines, and a $^{10}$B neutron capture layer, where each layer is separated by insulating SiO$_2$ layers.  The cross sectional image of the CB-KID.}
\end{figure}

The delay-line CB-KID was fabricated on a SiO$_2$/Si substrate. It has stacking layers of a superconducting Nb ground plane, $X$ and $Y$ superconducting Nb meanderlines, and a $^{10}$B neutron conversion layer as shown in Fig.~\ref{CBKID}.
Each layer was separated by an insulating SiO$_2$ layer.
The thickness of each layers are (1) 300\,nm for Nb ground plane, (2) 350\,nm for SiO$_2$ layer, (3) 50\,nm for $Y$ meanderline, (4) 150\,nm for SiO$_2$ layer, (5) 50\,nm for $X$ meanerline, (6) 150\,nm for SiO$_2$ layer, and (7) 70\,nm for $^{10}$B layer.   
The $X$ and $Y$ meanderlines of width 0.9\,$\mu$m and total length 151\,m were folded 10,000 times with 0.6\,$\mu$m spacing.
Therefore, the repetition pitch $p$ and length of each segment $h$ were 1.5\,$\mu$m and 15.1\,mm, respectively.
The device was fabricated in the clean room for analog--digital superconductivity (CRAVITY) at the National Institute of Advanced Industrial Science and Technology (AIST).
A $^{10}$B layer was deposited by electron-beam deposition under ultra-high vacuum.

We obtained the penetration depth $\lambda_{\rm s} \sim$ 115\,nm and the kinetic inductance of the meanderline $L_{\rm k} \sim 56\,\mu$H at 7.9\,K with the two-fluid model, where we take a superconducting transition temperature $T_{\rm c}$ = 8.5\,K and the penetration depth of Nb meanderlines $\lambda_{\rm s} =$ 60\,nm at 4.2K. 

The nuclear reaction between $^{10}$B atoms and incident neutrons at the neutron conversion layer creates charged particles, including ($^7$Li ions and $\alpha$-particles).
One of the particles could hit both the $X$ and $Y$ superconducting meanderlines, and thus, a transient reduction of $n_{\rm s}$ occurs locally at the hot spot.

According to the particle and heavy ion transport code system (PHITS) simulations, the energy deposition by $^7$Li-ions is comparable to that by $\alpha$-particles \cite{Mal20}. 
Therefore, one can expect a similar transient reduction in $n_{\rm s}$ regardless of whether a $^7$Li particle or an $\alpha$ particle is incident to a sensitive area of the CB-KID detector.

A local transient reduction of $n_{\rm s}$ induces a transient change in local kinetic inductance.
When a DC bias current $I_{\rm b}$ is passed through the meanderlines, a pair of voltage pulses are generated at the hot spot, and each pulse propagates toward both ends of the meanderline as an electromagnetic wave.
A voltage $V$ across the hot spot is expressed as
\begin{equation}
\label{eq:V}
V = I_{\rm b}\frac{{\rm d}L }{{\rm d}t}\simeq I_{\rm b}\frac{{\rm d}L_{\rm k}}{{\rm d}t} = -\frac{m_{\rm s}\Delta l I_{\rm b}}{n_{\rm s}^2 q_{\rm s}^2 S}\frac{{\rm d}n_{\rm s}}{{\rm d}t},
\end{equation}
where $L$ is inductance; $L_{\rm k}$ is kinetic inductance; $\Delta l$ and $S$ are the hot-spot length and cross-sectional area of the superconducting wire, respectively; and $m_{\rm s}$ and $q_{\rm s}$ are the effective mass and electric charge of the Cooper pair, respectively.

Meanderlines with the ground plane are regarded as superconducting microstriplines, and thus, they can transmit high-frequency waves with lower attenuation even for the 151\,m-length traveling \cite{Swirhart, Koy18}.
The meanderline acts as a superconducting detector and a delay line for the pulsed signals.
Therefore, one can determine the hot-spot positions on the detector with only four signal leads.
The hot-spot positions $x$ and $y$ are determined by
\begin{eqnarray}
\label{eq:X}
x={\rm ceil}\left[\frac{(t_{\rm Ch4}-t_{\rm Ch3})v_x}{2h}\right] p,
\end{eqnarray}
\begin{eqnarray}
\label{eq:Y}
y={\rm ceil}\left[\frac{(t_{\rm Ch2}-t_{\rm Ch1})v_y}{2h}\right] p,
\end{eqnarray}
where $t_{\rm Ch1}$, $t_{\rm Ch2}$, $t_{\rm Ch3}$, and $t_{\rm Ch4}$ are the corresponding timestamps of the signals received at Ch1, Ch2, Ch3, and Ch4, respectively, and $v_x = 5.966\times 10^7$\,m/s and $v_y = 5.469\times 10^7$\,m/s are signal propagation velocities for the $X$ and $Y$ meanderlines, respectively, at $T$ = 7.9\,K in this CB-KID.

We note that pixel sizes are not uniform by integerization processing in Eqs.~\ref{eq:X} and \ref{eq:Y}.  Pixel size in the $x$ direction contains majority of 3\,$\mu$m and a few of 1.5\,$\mu$m.  That in the $y$ direction consists of about the same number of 1.5\,$\mu$m and 3\,$\mu$m. 

If we assume events occurs at $t_0$, it is determined by
\begin{eqnarray}
t_0=\frac{t_{\rm Ch4}+t_{\rm Ch3}}{2}-\frac{l}{2v_x}, 
\label{t0:x}
\end{eqnarray}
or
\begin{eqnarray}
t_0=\frac{t_{\rm Ch2}+t_{\rm Ch1}}{2}-\frac{l}{2v_y}.
\label{t0:y}
\end{eqnarray}
These $t_0$ should be the same if the signal quartet originated from the same event. 
This coincidence check act as a good criterion to distinguish the correct quartets from the bundle of signals.
In other word, the delay-line CB-KID has a multihit tolerance.

The structure and principle of the detector are described in detail in \cite{Iiz19}.

\subsection{Preparation of GdAl$_3$ single crystals in a Gd--Al alloy ingot}
GdAl$_3$ is an antiferromagnet with the N\'eel temperature of 18\,K \cite{Bus66}.
Herein, single crystals of GdAl$_3$ were grown using the Al self-flux method.
The starting materials, Gd and Al with an atomic ratio of 1:9, were sealed in a quartz tube under Ar atmosphere.
The tube was heated to 1000\,$^\circ$C and cooled to 650\,$^\circ$C at a cooling rate of 10\,$^\circ$C/h.
Ingots of Gd 5\% Al 95\% eutectic alloy containing GdAl$_3$ single crystals were obtained.
Platelet specimens of thickness $\sim$1\,mm were cut from the ingots for the neutron-imaging experiments.

\subsection{Cryogenic and signal measurement systems}
The delay-line CB-KID was cooled to 7.9\,K with Gd--Al alloy sample using a Gifford--McMahon (GM) refrigerator in which the cold head is suspended by silicone rubber feet for vibration isolation.
A test sample was pasted on an Al plate using an epoxy adhesive, which was placed at a distance of 0.8\,mm from the detector.
The temperature of the detector was modulated using a temperature controller (Model 44 by Cryogenic Control Systems Inc.).
The vacuum can and constituent parts of the sample holder were made using Al alloy, and the alloy is almost transparent to neutron beams.

The signals from four channels were amplified by an ultralow-noise differential amplifier (SA-430 F5 by NF Corporation), and the negative signals (Ch1 and Ch3) were inverted.
Therefore, the Kalliope-DC readout circuit, a 1\,ns-sampling multichannel (16\,Ch\,$\times$\,2) time-to-digital converter \cite{Shi18}, and a 2.5-GHz sampling digital oscilloscope (Teledyne LeCroy HDO4104-MS) simultaneously received a quartet of positive polarity signals.
A signal arrival timestamp was recorded by Kalliope-DC circuit when a signal intensity of each channel reached preset threshold, which was tuned at slightly above noise levels. Thus, most of signal timestamps were detected at the arising edge of the signals.
Both oscilloscope and Kalliope-DC circuit were triggered to be measured in synchronization with pulsed neutron beams.

\begin{figure}[h]
\begin{center}
\includegraphics[width=7cm, pagebox=cropbox, clip]{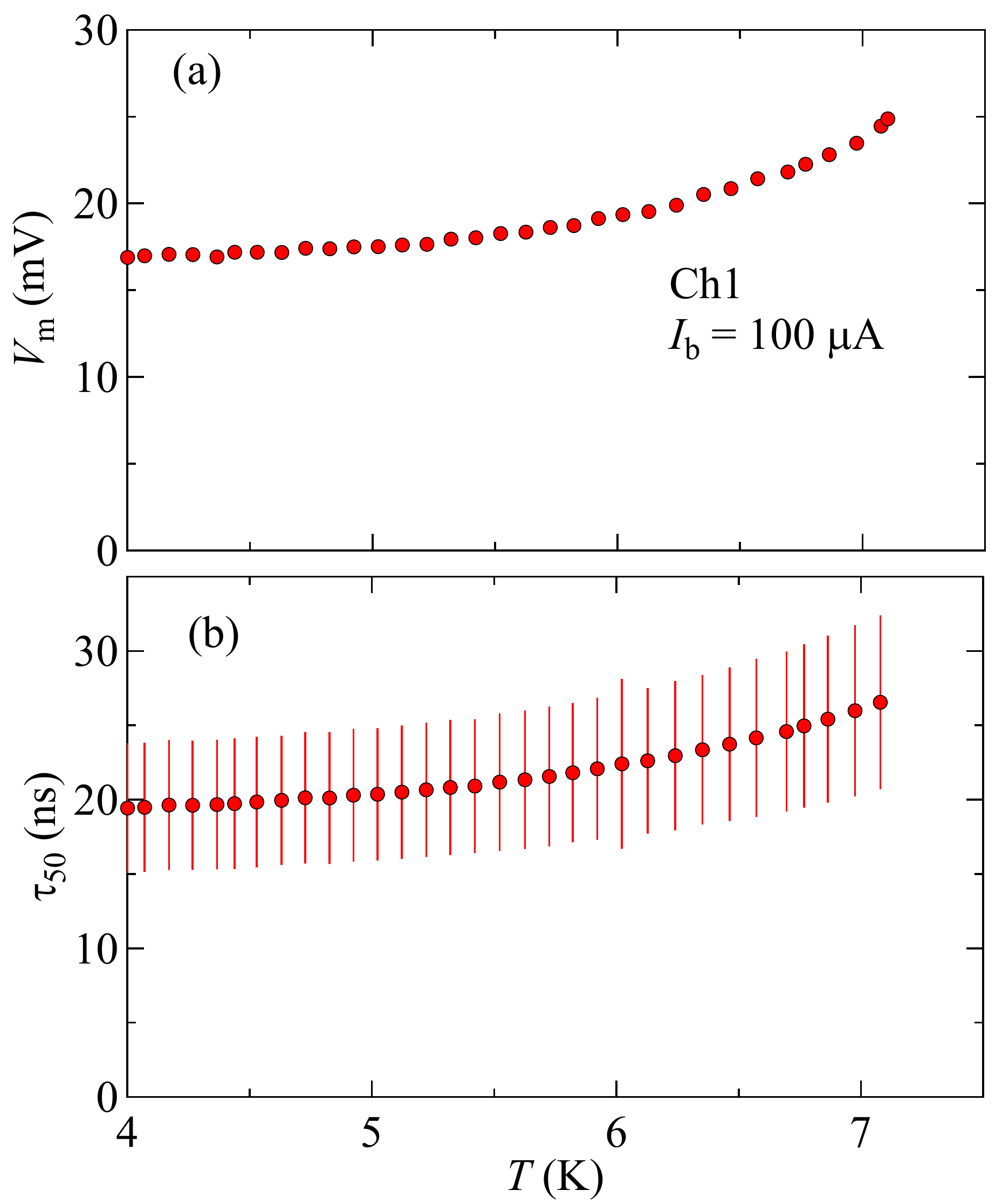}
\end{center}
\caption{\label{signal}Temperature dependence of (a) the median of signal intensities $V_{\rm m}$ and (b) average of the full width at half maximum of signals $\tau_{50}$---the error bars represent the standard deviation.}
\end{figure}

\subsection{Neutron imaging}
Neutron-imaging experiments were performed using pulsed neutrons at BL10 of the Material and Life Science experimental Facility (MLF), J-PARC \cite{BL10}.
Neutrons traveled from the moderator to the detector through the 14\,m beamline, where TOF ($t$) was proportional to the neutron wavelength $\lambda$ ($\lambda$ (nm) = 28.2556 \,$\times t$ (sec)) at BL10.
Neutron energy $E$ is expressed as $E$ (meV)=0.8179/$\lambda^2$.
The uncertainty of TOF was 33 \,$\mu$s in the full width at half maximum (FWHM) for 10\,meV neutrons.
The collimator ratio of $L/D$=140 was used for the imaging experiments.
Neutron beams were irradiated to the detector from the substrate side through the test sample.

\section{Results and discussion}

\subsection{Temperature variation of signals}
The main advantage of a CB-KID is the wide operating range.
It can operate in a wide bias current range and a wide temperature range within the superconducting phase.
Variation of the operating temperature of the detector with the median of the signal intensities $V_{\rm m}$ at $I_{\rm b}=100\,\mu$A with a beam power of 307 \,kW is shown in Fig.~\ref{signal}(a).
$V_{\rm m}$ increases gradually with an increase in temperature and increases rapidly toward $T_{\rm c}$ above 6 \,K.
This temperature dependence can be qualitatively clarified by the reduction of $n_{\rm s}$ as $T_{\rm c}$ approaches.

Figure~\ref{signal}(b) shows the operating-temperature dependence of the average FWHM of signals $\tau_{50}$ that gradually increases with an increase in temperature, but its increasing rate is lower than that of $V_{\rm m}$.
Assuming that detection-rate tolerance is in the order of 1/$\tau_{50}$ to separate closest signals, a deterioration in the detection-rate tolerance due to temperature increase becomes moderate. Detection-rate tolerance at 7\,K at a maximum is expected to be around 30\,MHz which is 75\% of that at 4\,K.

Reduction of $v_x$ and $v_y$ at higher operating temperatures \cite{Vu20} reduces the pixel size of the neutron transmission image, as shown in Eqs.~\ref{eq:X} and \ref{eq:Y}.
Considering the improvement in detection efficiency by increasing the operating temperature of the detector \cite{Vu20}, operating CB-KID at high temperatures improves imaging.

\begin{figure}[t]
\begin{center}
\includegraphics[width=7cm, pagebox=cropbox, clip]{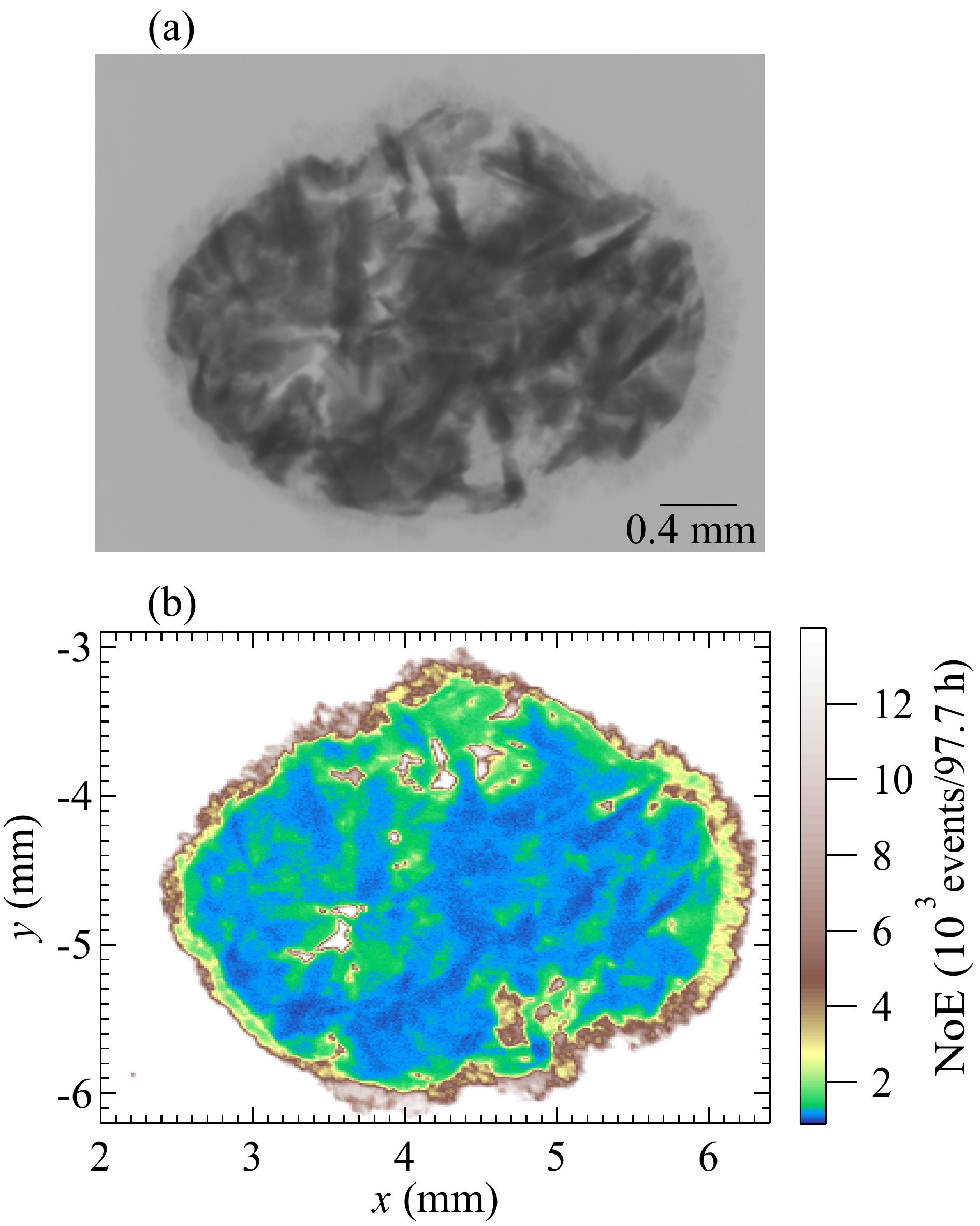}
\end{center}
\caption{\label{images}(a) X-ray and (b) neutron transmission images---the shape of the sample and the position of single crystals of GdAl$_3$ are identified.}
\end{figure}

\begin{figure}[h]
\begin{center}
\includegraphics[width=7cm, pagebox=cropbox, clip]{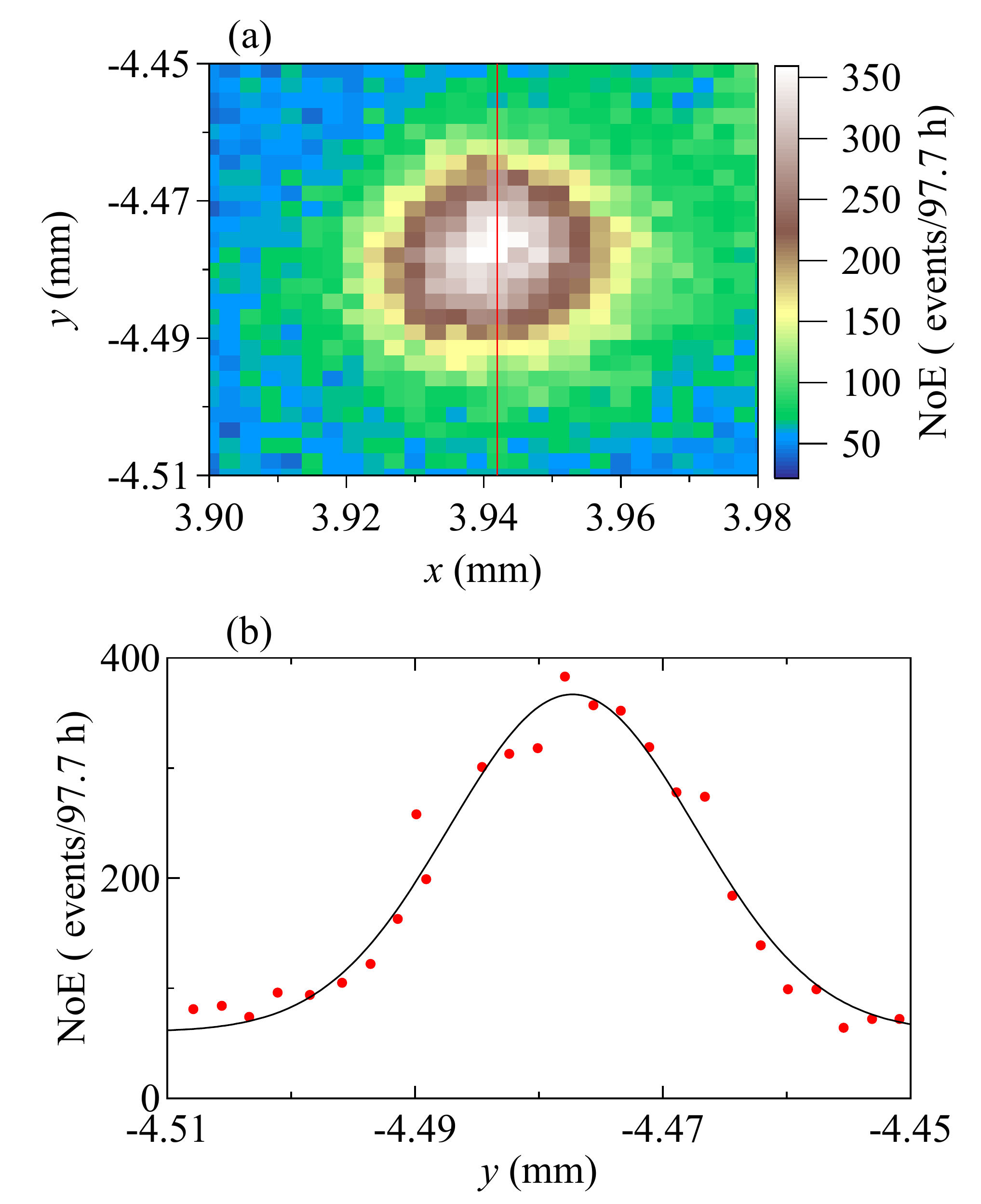}
\end{center}
\caption{\label{Prof}(a) Enlarged image near a hole (see Fig.~\ref{images}(b)), and (b) line profile along the red solid line in (a)---the black solid line represents the Gaussian fitting.}
\end{figure}

\subsection{Neutron transmission image}

Figure~\ref{images}(a) shows the X-ray transmission image of the Gd--Al alloy sample.
The dark regions with rectangles represent GdAl$_3$ single crystals, in which Gd atoms have huge X-ray total cross sections because of the large atomic number.
There are some bright regions that correspond to holes or Al-atom aggregated areas.

The neutron transmission image of the sample is shown in Fig,~\ref{images}(b).
The number of events (NoE) with an incident neutron wavelength $\lambda$ ranging from 0.052 to 1.13\,nm were combined up to 25\,pixels ---5\,pixels for $x$ and $y$ directions each--- to obtain a high-contrast image, and it was represented using a color scale.
Accumulation time of events was 97.7\,h for 517\,kW of beam power under the detector conditions of $T=7.9\,$K and $I_{\rm b}^x = I_{\rm b}^y = $15\,$\mu$A.
GdAl$_3$ single crystals are well identified as rectangular-shaped dark-blue regions, where neutrons are well absorbed by Gd nuclei.
The shapes and arrangements of GdAl$_3$ single crystals in the neutron and X-ray transmission images are consistent.
Neutron transmission was low throughout the sample as Gd nuclei were distributed in the sample as an Al-rich Al--Gd alloy.
Because the Gd nuclei have the largest neutron absorption cross-section with $\lambda \geq$ 0.1\,nm, Gd nuclei are detected even at the low Gd concentrations (see the green-colored regions of Fig.~\ref{images}(b).)

\begin{figure}[h]
\begin{center}
\includegraphics[width=7cm, pagebox=cropbox, clip]{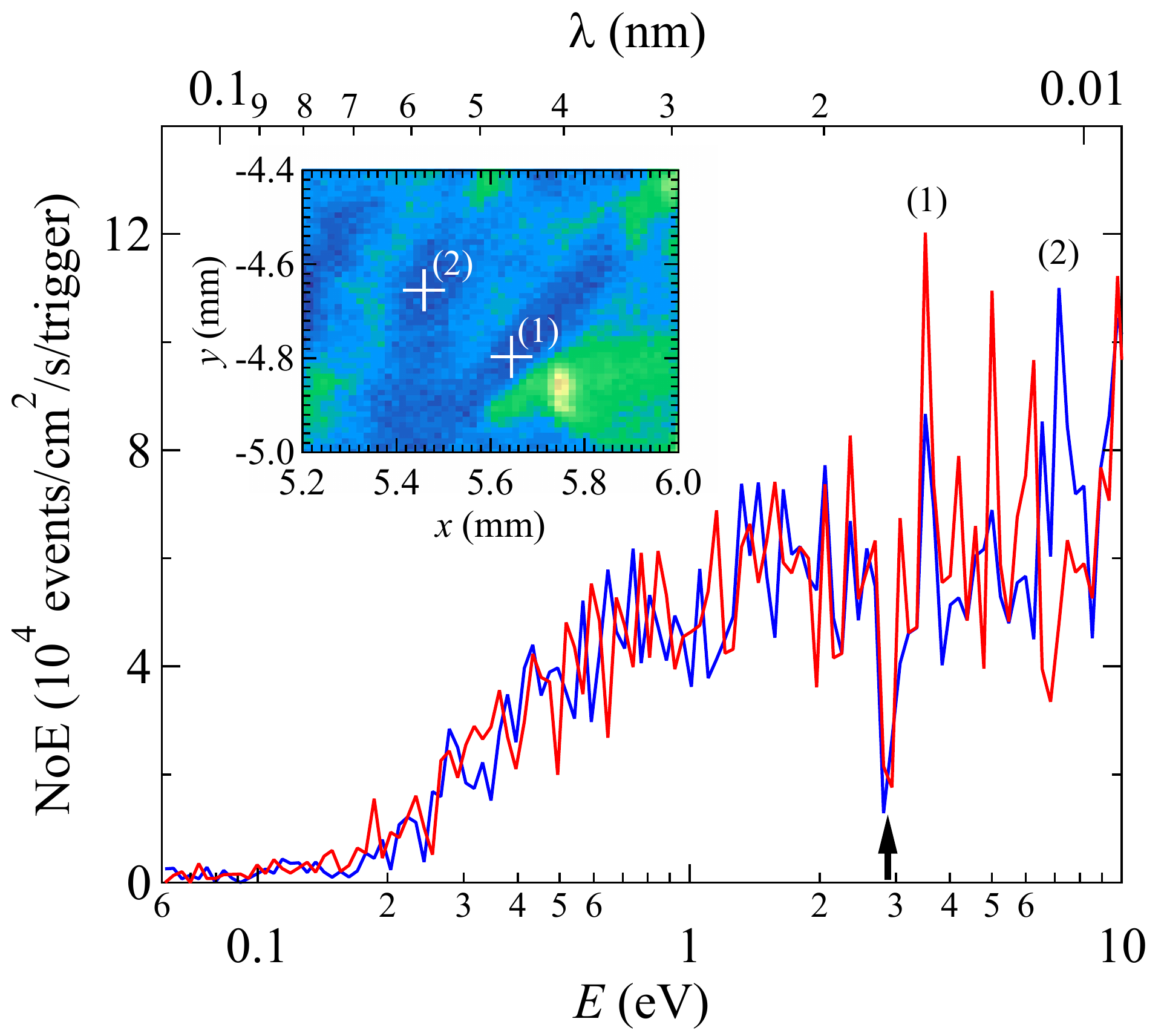}
\end{center}
\caption{\label{res_dip} Neutron wavelength $\lambda$ dependence of the number of events (NoE) for GdAl$_3$ single crystals at crosses on the inset that shows an enlarged image of Fig.~\ref{images}(b).}
\end{figure}

\begin{figure}
\begin{center}
\includegraphics[width=7cm, pagebox=cropbox, clip]{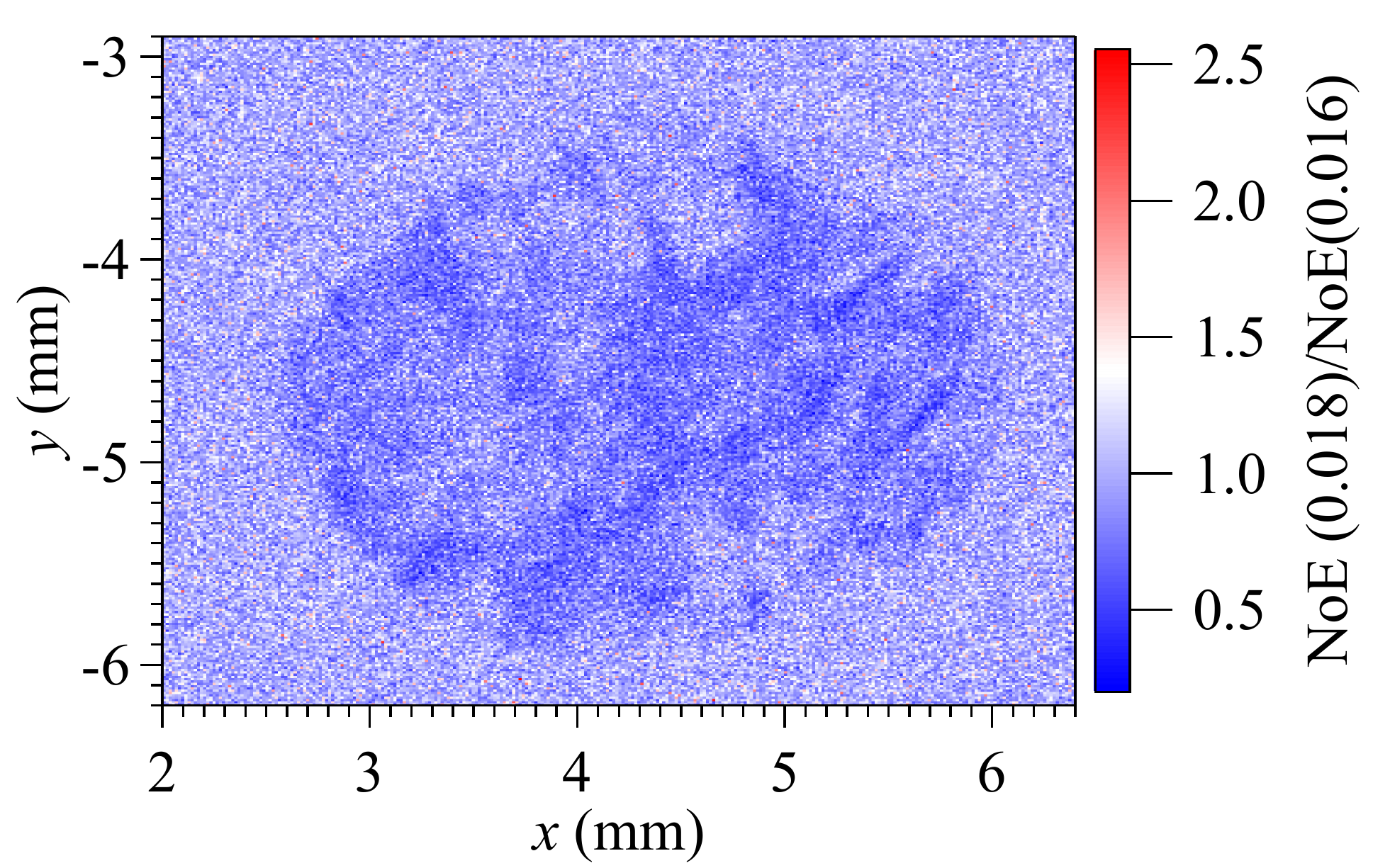}
\end{center}
\caption{\label{NRTI} Gd selective image obtained by dividing the neutron transmission image with a wavelength at the resonance dip of $^{155}$Gd of 0.018\,nm (2.5\,eV) bin by that with a wavelength of 0.016\,nm (3.2\,eV) bin.}
\end{figure}

Figure~\ref{Prof}(a) shows the zoomed image of a hole in Fig.~\ref{images}(b), and the NoE is computed by summing up for the area of the $1\times1$ pixel.
The line profile corresponding to the red line in Fig.~\ref{Prof}(a) is fitted using a Gaussian function as shown in Fig.~\ref{Prof}(b).
The FWHM of the hole size was 27.7\,$\pm$0.9\, and 23.4\,$\pm1.2\,\mu$m for the $x$ and $y$ directions, respectively.
Therefore, we conclude that CB-KID with the operating temperature of 7.9\,K can distinguish at least a 25\,$\mu$m-diameter hole.

\subsection{Neutron resonance transmission imaging}

A zoomed image of Fig.~\ref{images}(b) is shown in the inset of Fig.~\ref{res_dip}.
The neutron energy $E$ or wavelength $\lambda$ dependence of NoE for GdAl$_3$ single crystals at white crosses positions are shown in the main panel of Fig.~\ref{res_dip}.
Two curves exhibit almost the same energy dependence even though they are associated with the different small crystals of the sample.
Herein, the NoE was computed by summing up for the area of 15 $\times$ 12 $\mu$m$^2$ at each energy bin, and a characteristic NoE distribution for Gd nuclei was observed even for a limited area.
NoE is reduced to almost zero above $E <$ 0.2\,eV ($\lambda >$ 0.07\,nm).
The $\lambda$ dependence is well reproduced by increasing the total neutron cross sections for $^{155}$Gd and $^{157}$Gd nuclei above $E <$ 0.9\,eV ($\lambda >$ 0.03\,nm).
The resonance dip shown using the arrow corresponds to the resonance peak of $^{155}$Gd nuclei at $E$ = 2.58\,eV  \cite{JENDL}.

Figure~\ref{NRTI} shows the neutron transmission image obtained by dividing the neutron transmission image with a wavelength at the resonance dip of 0.018\,nm (2.5\,eV) bin by that with a wavelength of 0.016\,nm (3.2\,eV) bin.
Herein, NoEs from $5\times5$ pixels were binned over a wavelength range of 0.001\,nm to obtain the neutron transmission images at $\lambda$ = 0.018 and 0.016\,nm.
At a resonance dip of $^{155}$Gd, GdAl$_3$ single crystals are clearly observed even in a narrow wavelength range.
It demonstrates the capability for NRTI with a high spatial resolution for the delay-line CB-KID.

\section{Conclusion}
We performed high spatial resolution neutron transmission imaging for a Gd--Al alloy sample using a delay-line CB-KID at an operating temperature of 7.9\,K.
GdAl$_3$ single crystals distributed throughout the Gd--Al alloy were observed by neutron transmission imaging. 
It may pave the way for the delay-line CB-KID to conduct high spatial imaging of crystal growth from the liquid phase without destroying the samples.    
Gd atoms were identified by the characteristic suppression of the transmission by increasing the neutron capture cross section of $^{155}$Gd and $^{157}$Gd.
Furthermore, the resonance dip of $^{155}$Gd was visible even when the NoE was in the limited area of 15 $\times$ 12 $\mu$m$^2$.
We further performed the Gd NRTI using the $^{155}$Gd resonance dip.

\section{ACKNOWLEDGMENT}
This work is partially supported by Grant-in-Aid for Scientific Research (Nos. JP16H02450, JP21H04666, JP21K14566) from JSPS. The neutron irradiation experiments at the Materials and Life Science Experimental Facility (MLF) of the J-PARC were conducted under the support of MLF programs (No. 2020P0201).

\end{document}